\documentclass[12pt]{article}

\usepackage{graphicx}
\usepackage[fleqn]{amsmath}
\usepackage{amssymb,latexsym,multirow}
\usepackage{psfrag}

\begin{document}

\title{Quasi-exact solutions of nonlinear differential equations}

\author{Nikolay A. Kudryashov, Mark B. Kochanov}

\date{Department of Applied Mathematics, National Research Nuclear University
MEPHI, 31 Kashirskoe Shosse, 115409 Moscow, Russian Federation}

\maketitle

\begin{abstract}
The concept of quasi-exact solutions of nonlinear differential equations is introduced.
Quasi-exact solution expands the idea of exact solution for additional values of parameters of differential equation.
These solutions are approximate solutions of nonlinear differential equations but they are close to exact solutions.
Quasi-exact solutions of the the Kuramoto--Sivashinsky, the Korteweg--de Vries--Burgers and the Kawahara equations are founded.
\end{abstract}

\section{Introduction}

During the past years we have observed the intensive study of nonlinear DEs.
Many methods have been introduced for finding exact solutions of nonlinear DEs.
Such methods include the singular manifold method \cite{WTC, Kudryashov88, Conte89, Kudryashov90, Kudryashov90a, Kudryashov91}, the tanh-expansion method \cite{Malfliet, Hereman01, Parkes01, Hereman02, Hereman03,  Kudr96a}, the simplest equation method \cite{Kudr05, Kudr08, Kudr10, Vitanov01, Vitanov02, Vitanov03, Biswas}, the $G'/G$-expansion method \cite{Wang, Kudr09c, Biswas01, Kudr10c} and the direct method for finding all exact solutions \cite{Kudr10a, Kudr10b, Kudr11, Kudr11aa, Kudr12aa}.
However applying such methods for constructing exact solutions of nonlinear DEs that arise in different mathematical models we have to take into account that most mathematical models do not describe physical processes exactly.
Often all experimental and computational techniques are introduce some errors.
This  means that we do not deal with exact mathematical models  describing physical processes.
So, exact solutions of nonlinear DEs are not necessary better than approximate solutions.

In this paper we build on the concept of exact solutions for nonlinear DEs and introduce by introducing a definition of quasi-exact solutions for nonlinear DEs.
These solutions are approximate ones which are close to exact solutions.
When we deal with integrable nonlinear DEs we do not need  quasi-exact solutions because  we can find  exact solutions for all values of the parameters.
Our approach can be used for nonintegrable nonlinear DEs when we can obtain some exact solutions.

Let us introduce the definition of the quasi-exact solutions for nonlinear DEs. Assume we have the nonlinear ODE in the form
\begin{equation}
  \begin{gathered}
  \label{Eq0}
  E(y, y_z, y_{zz}, \dotsc, z, \alpha_1, \dotsc, \alpha_n)=0,
  \end{gathered}
\end{equation}
where $z$ is the independent variable, $y(z)$ is a solution of Eq.\eqref{Eq0}, and $\alpha_i$ ($i=1, \dotsc, n$) are some constant real parameters.

Let $y^*(z, \alpha_1^*, \dotsc, \alpha_n^*)$ be an limited exact solution of Eq.\eqref{Eq0} at all real z such that
\begin{equation}
  \begin{gathered}
  \label{Eq0a}
  E(y^*, y^*_z, y^*_{zz}, \dotsc, z, \alpha_1^*, \dotsc, \alpha_n^*) \equiv 0.
  \end{gathered}
\end{equation}

\emph{\textbf{Definition.}
The quasi-exact solution of Eq.\eqref{Eq0} is called an dependence with respect to $z$: $y=y^*(z, \alpha_1, \dotsc, \alpha_n)$, $(\alpha_1, \dotsc, \alpha_n) \subset \Omega \subset \mathbb{R}^n$ of \eqref{Eq0}, where $\Omega$ is such that
\begin{equation}\label{Eq0b}
\underset{z}{\textnormal{max}} |E(y^*,y^*_z,y^*_{zz}, \dotsc, \alpha_1, \dotsc, \alpha_n)| < \varepsilon, \quad (\alpha_1, \dotsc, \alpha_n) \subset \Omega
\end{equation}
and $\varepsilon$ is a small quantity.}

Let us denote deviation function
\begin{equation*}
R(z; \alpha_1, \dotsc, \alpha_n) \equiv E(y^*,y^*_z,y^*_{zz}, \dotsc, \alpha_1, \dotsc, \alpha_n)
\end{equation*}
with norm
\begin{equation*}
\|R(\alpha_1, \dotsc, \alpha_n)\| = \underset{z}{\text{max}} |R(z; \alpha_1, \dotsc, \alpha_n)|.
\end{equation*}

An essence, the substituting of the quasi-exact solutions into equation does not yield zero but a  small quantity.
However  we have some advantages in comparison with exact solutions because the quasi-exact solution allows us to have approximate solution in the case of other values for the parameters of nonlinear DEs.

\section{Method applied}

The quasi-exact solution of nonlinear DE can be found if we solve the following equation with respect to the parameters $\alpha_1, \dotsc, \alpha_n$
\begin{equation}\label{Eq0bb}
\underset{z}{\text{max}} |E(y^*,y^*_z,y^*_{zz}, \dotsc, \alpha_1, \dotsc, \alpha_n)| - \varepsilon=0, \quad (\alpha_1, \dotsc, \alpha_n) \subset \Omega,
\end{equation}
where $\varepsilon$ is the given small quantity, $w^*(z)$ is the expression of exact solution with unknown parameters $\alpha_1, \dotsc, \alpha_n$ of Eq.\eqref{Eq0}.

The algorithm for finding the quasi-exact solutions of nonlinear evolution DEs is based on the method presented in paper \cite{Kudr12b}.
This approach contains the following steps.

\emph{Step 1: Reduction of a nonlinear evolution equation to an ODE.}

In the case of nonlinear PDE in polynomial form,
\begin{equation}
  \begin{gathered}
  \label{Eq10a}
  E_1(u, u_t, u_x, u_{tt}, u_{xx}, \dotsc, \alpha_1, \dotsc, \alpha_n) = 0,
      \end{gathered}
\end{equation}
we reduce the nonlinear PDE to a nonlinear ODE by seeking traveling wave solutions
\[u(x,t)=y(z), \qquad z=k\,x-\omega\,t.\]

Result is a nonlinear ODE with parameters $\omega$ and $k$ in  the form
\begin{equation}
  \begin{gathered}
  \label{Eq10b}
  E_2(y, \omega y_z,\, k y_z,\, \omega^2 y_{zz},\, k^2 y_{zz}, \dotsc, \alpha_1, \dotsc, \alpha_n)=0.
      \end{gathered}
\end{equation}

\emph{Step 2: Finding solution of Eq.\eqref{Eq10b} of the form}
\begin{equation}
  \begin{gathered}
  \label{Eq10c}
  y(z)=\sum_{n=0}^{N}\,a_n\,Q^n,
      \end{gathered}
\end{equation}
where $N$ is the pole order of the general solution for Eq.\eqref{Eq10b} and $Q(z)$ is of the form
\begin{equation}
  \begin{gathered}
  \label{Eq10d}
 Q(z)=\frac{1}{1+e^{z}}.
      \end{gathered}
\end{equation}

\emph{Remark 2.1.} The value $N$ in expression \eqref{Eq10c} is determined by substituting $y(z)=a_0\,z^{-N}$, where $N>0$ into all monomials of Eq.\eqref{Eq10b} and comparing two or more terms with the smallest powers in the equation.
We can continue with our method if $N$ is an integer.
If $N$ is noninteger we need to use a transformation of solution $y(z)$.
For example, if we obtain the value $N=\frac{1}{m}$, where $m$ is an integer we can use the transformation for the solution in the form $y(z)=w(z)^\frac{1}{m}$, where $w(z)$ is the new dependent variable.

We note that the function $Q$ is the solution of the equation
\begin{equation}
  \begin{gathered}
  \label{Eq10e}
 Q_z=Q^2-Q.
      \end{gathered}
\end{equation}

\emph{Step 3: Substitution of $y(z)$ and its $z$-derivatives into Eq.\eqref{Eq10b}.}

As a result of the third step we obtain the equation involving  the function  $Q$, coefficients $a_n$, where  $(n=0,1, \ldots, N)$ and the parameters $\omega$, $k$ of Eq.\eqref{Eq10b}.

\emph{Step 4: Finding algebraic equations for the coefficients $a_n$ of formula \eqref{Eq10c} and equations for the parameters of Eq.\eqref{Eq10b}.}

Equating expressions at the different powers of $Q$ to zero we obtain the system of algebraic equations in the form
\begin{equation}
  \begin{gathered}
  \label{Eq10f}
      P_i(a_N, a_{N-1}, \ldots, a_0, k, \omega, \alpha_1, \dotsc, \alpha_n)=0,\qquad (i=0,\ldots, M).
      \end{gathered}
\end{equation}

\emph{Step 5: Finding values of the coefficients $a_n$ $(n=1,\ldots   N)$ for formula \eqref{Eq10c}.}

At this step  we obtain values of the coefficients $a_N$, $a_{N-1}$, ..., $a_0$ solving some equations of the system \eqref{Eq10f}.

\emph{Step 6: Substituting expression \eqref{Eq10c} into Eq.\eqref{Eq10b} and using the numerical method we solve equation \eqref{Eq0bb} and find values of parameters of  solution for Eq.\eqref{Eq10b}.}

Doing so, we obtain the values of parameters for the approximate solutions of nonlinear DE \eqref{Eq10b}. It is important to note the definition of quasi-exact solutions allows us to have approximate solution of nonlinear DE even if we can not solve the system of equations \eqref{Eq10f} exactly.

\section{Quasi-exact solutions of the Kuramoto--Sivashinsky equation}

Let us find the quasi-exact solutions of the Kuramoto--Sivashinsky equation
\begin{equation}
  \begin{gathered}
  \label{KurSivEq}
  v_t+v\,v_x+\alpha\,v_{xx}+\beta\,v_{xxx}+\gamma\,v_{xxxx}=0,
      \end{gathered}
\end{equation}
where $\alpha$, $\beta$ and $\gamma$ are parameters.

Nonlinear evolution equation \eqref{KurSivEq} has been studied by a number of authors from various viewpoints.
This equation has drawn much attention not only because it is interesting as a simple one-dimensional nonlinear evolution equation including effects of instability and dissipation but also it is important for description of engineering and scientific problems.
Equation \eqref{KurSivEq} was used in \cite{Kuramoto76} to explain the origin of persistent wave propagation through medium of reaction-diffusion type.
In \cite{Sivashinsky83}, equation \eqref{KurSivEq} was derived for the description of the nonlinear evolution of the disturbed flame front.
One also encounters \eqref{KurSivEq} for studying of motion of a viscous incompressible fluid flowing down an inclined plane \cite{Benney66, Topper78, Shkadov77}.
Mathematical modeling of dissipative waves in plasma physics by means of equation \eqref{KurSivEq} was presented in \cite{Cohen76}.
Elementary particles as the solutions of the Kuramoto--Sivashinsky equation were studied in \cite{Michelson90}.
Equation \eqref{KurSivEq} also can be used for description of nonlinear long waves in a viscous-elastic tube \cite{Kudryashov08a}.

Exact solutions of the Kuramoto--Sivashinsky equation are well known.
The solutions of Eq.\eqref{KurSivEq} were first found  by Kuramoto \cite{Kuramoto76} at $\beta=0$.
In the case $\beta\neq 0$ exact solutions of Eq.\eqref{KurSivEq} were first obtained in papers \cite{Kudryashov88, Kudryashov90}.
Later Eq.\eqref{KurSivEq} and its generalizations were considered many times.
The exact solutions of this equation also were re-discovered in \cite{Conte89, Kudryashov90a, Kudryashov91,  Zhu96, Berloff97, Fu05, Zhang06, Khuri07, Nickel07, Kudryashov07, Qin08, Kudr08bb}.

Using variables
\begin{equation}
  \begin{gathered}
  \label{Eq1a}
v=\frac{\alpha\,\sqrt{\alpha}}{\sqrt{\gamma}}\,u, \quad x=\sqrt{\frac{\gamma}{\alpha}}\,x^{'},\quad t=\frac{\gamma}{\alpha^2}\,t^{'}\quad \sigma=\frac{\beta}{\sqrt{\alpha\,\gamma}},
      \end{gathered}
\end{equation}
we obtain the Kuramoto--Sivashinsky equation in the form
\begin{equation}
  \begin{gathered}
  \label{Eq1b}
  u_t+u\,u_x+\,u_{xx}+\sigma\,u_{xxx}+\,u_{xxxx}=0.
      \end{gathered}
\end{equation}

Seeking traveling wave solutions
\begin{equation}
  \begin{gathered}
  \label{Eq1d0}
u(x,t)=y(z), \quad z=k\,x-\omega\,t,
      \end{gathered}
\end{equation}
after one integration with respect to $z$, we obtain the nonlinear ODE of third order,
\begin{equation}
  \begin{gathered}
  \label{Eq11dd}
 k^4\,y_{zzz} +k^3\sigma\,y_{zz}+k^2\,y_{z}+\frac{k}{2}\,y^2-\omega\,y + C_1 = 0, \quad k \neq 0.
      \end{gathered}
\end{equation}

Using a new variable $y(z)=y^{'}(z)+\frac{C_0}{k}$ (the primes are omitted below) we transform Eq.\eqref{Eq11dd} into the form
\begin{equation}
  \begin{gathered}
  \label{Eq1d}
 k^4\,y_{zzz} +k^3\sigma\,y_{zz}+k^2\,y_{z}+\frac{k}{2}\,y^2-\frac{\omega^2}{2\,k} + C_1=0, \quad k \neq 0.
      \end{gathered}
\end{equation}

Eq.\eqref{Eq1d} is invariant under transformations $y\rightarrow -y$, $z\rightarrow -z$,  $\sigma \rightarrow -\sigma$.
Thus, without loss of generality we consider the case $\sigma\geq 0$.
Eq.\eqref{Eq1d} may possess solutions with third-order poles and we look for the quasi-exact solution in the following form
\begin{equation}
  \begin{gathered}
  \label{Eq1f}
y=a_0+a_1\,Q+a_2\,Q^2+a_3\,Q^3.
      \end{gathered}
\end{equation}
Substituting \eqref{Eq1f} into Eq.\eqref{Eq1d} and equating to zero the different power of $Q(z)$ we obtain the system
\begin{gather}
Q^6: 120 k^3 + a_3 = 0, \\
Q^5: a_2 a_3  +  24 k^3 a_2 - 144 k^3 a_3 + 12 k^2 \sigma a_3 = 0, \\
Q^4: 12 k^2 \sigma a_2 - 42 k^2 \sigma a_3 + a_2^2 + 2 a_1 a_3 - 108 k^3 a_2
+ 12 k^3 a_1 + 222 k^3 a_3 + 6 k a_3 = 0,
\\
\begin{split}
Q^3: - 2 k^2 \sigma a_1 &+ 3 k a_3 + 10 k^2 \sigma a_2 - 9 k^2 \sigma a_3 - 38 k^3 a_2 \\
&- 2 k a_2+ 27 k^3 a_3 - a_1 a_2 + 12 k^3 a_1 - a_0 a_3 = 0,
\end{split} \\
Q^2: 14 k^3 a_1 + 8 k^2 \sigma a_2 + 2 a_0 a_2 - 6 k^2 \sigma a_1 + a_1^2 + 2 k a_1 - 16 k^3 a_2 - 4 k a_2 = 0, \\
Q^1: a_1 ( - k^2 \sigma - a_0 + k^3 + k ) = 0, \\
Q^0: 2 C_1 k + k^2 a_0^2 - \omega^2 = 0.
\end{gather}
Solving this system, we have
\begin{equation}
\label{Eq1k}
\begin{split}
&a_3 = -120\,k^3, \quad
a_2 = 15\,{k}^{2} \left( 12\,k-\sigma \right), \quad
a_1 = \frac{15}{76} k \sigma^2-\frac{60}{19} k - 60 k^3  + 15 k^2 \sigma, \\
&a_0 = - \frac{5}{4} k^2 \sigma - \frac{15}{152} k \sigma^2 + \frac{30}{19} k - \frac{13}{608} \sigma^3 + \frac{7}{76} \sigma, \quad
C_1 =\frac{\omega^2}{2\,k}-\frac{k a_0^2}{2}, \\
&k_{1,2} = \pm \frac{1}{76} \sqrt{1520 - 95\sigma^2 + 19 \sqrt{549\sigma^4 - 3584\sigma^2 + 9216}}, \\
&k_{3,4} = \pm \frac{1}{76} \sqrt{1520 - 95\sigma^2 - 19 \sqrt{549\sigma^4 - 3584\sigma^2 + 9216}}.
\end{split}
\end{equation}

In addition we obtain the equation at $Q(z)^1$:
\begin{equation}
( -76 k \sigma - \sigma^2 + 304 k^2 + 16 ) ( 152 k^2 \sigma - 352 k - 56 \sigma + 60 k \sigma^2 + 13 \sigma^3 + 608 k^3 ) = 0,
\end{equation}
where $k=k_{1,2,3,4}$.
This equation has exact solutions
\begin{equation*}
\begin{aligned}
&k=k_{1,2}, \quad \sigma_\text{exact} = 0,4, \\
&k=k_{3,4}, \quad \sigma_\text{exact} = \frac{12}{\sqrt{47}}, \frac{16}{\sqrt{73}}.
\end{aligned}
\end{equation*}
This result was first obtained in \cite{Kudryashov88, Kudryashov90}.

Substituting the expression
\begin{multline}
\label{eq:kuramoto_sol_a}
y(z;\sigma)= - \frac{5}{4} k^2 \sigma - \frac{15}{152} k \sigma^2 + \frac{30}{19} k - \frac{13}{608} \sigma^3 + \frac{7}{76} \sigma + \\
+\left(\frac{15}{76} k \sigma^2-\frac{60}{19} k - 60 k^3  + 15 k^2 \sigma\right)\,Q(z)
+15\,{k}^{2} \left( 12\,k-\sigma \right)\,Q(z)^2 -\\
-120\,k^3\,Q(z)^3
\end{multline}
into Eq.\eqref{Eq11dd} at the condition $C_1=\frac{\omega^2}{2\,k}-\frac{k a_0^2}{2}$ we obtain
\begin{multline}
R(z;\sigma) = \frac{15k^2}{46208} ( 16 - 76 k \sigma - \sigma^2 + 304 k^2) \\
( 152 k^2 \sigma - 352 k - 56 \sigma + 60 k \sigma^2 + 13 \sigma^3 + 608 k^3 ) Q(z).
\end{multline}
Note that the maximum of the function $Q(z)$ is equal to 1.
Taking into account this fact and maximizing the function $R(z;\sigma)$ with respect to variable $z$ we get the dependence $\|R\|$ on $\sigma$ in the form
\begin{multline}
\|R(\sigma)\| = \frac{15k^2}{46208} |(16 - 76 k \sigma - \sigma^2 + 304 k^2) \\
( 152 k^2 \sigma - 352 k - 56 \sigma + 60 k \sigma^2 + 13 \sigma^3 + 608 k^3 )|.
\label{eq:R_sigma}
\end{multline}

\begin{figure}
\psfrag{xlbl}{$\sigma$}
\psfrag{ylbl}{$\|R(\sigma)\|$}
\center{\includegraphics[width=0.7\linewidth]{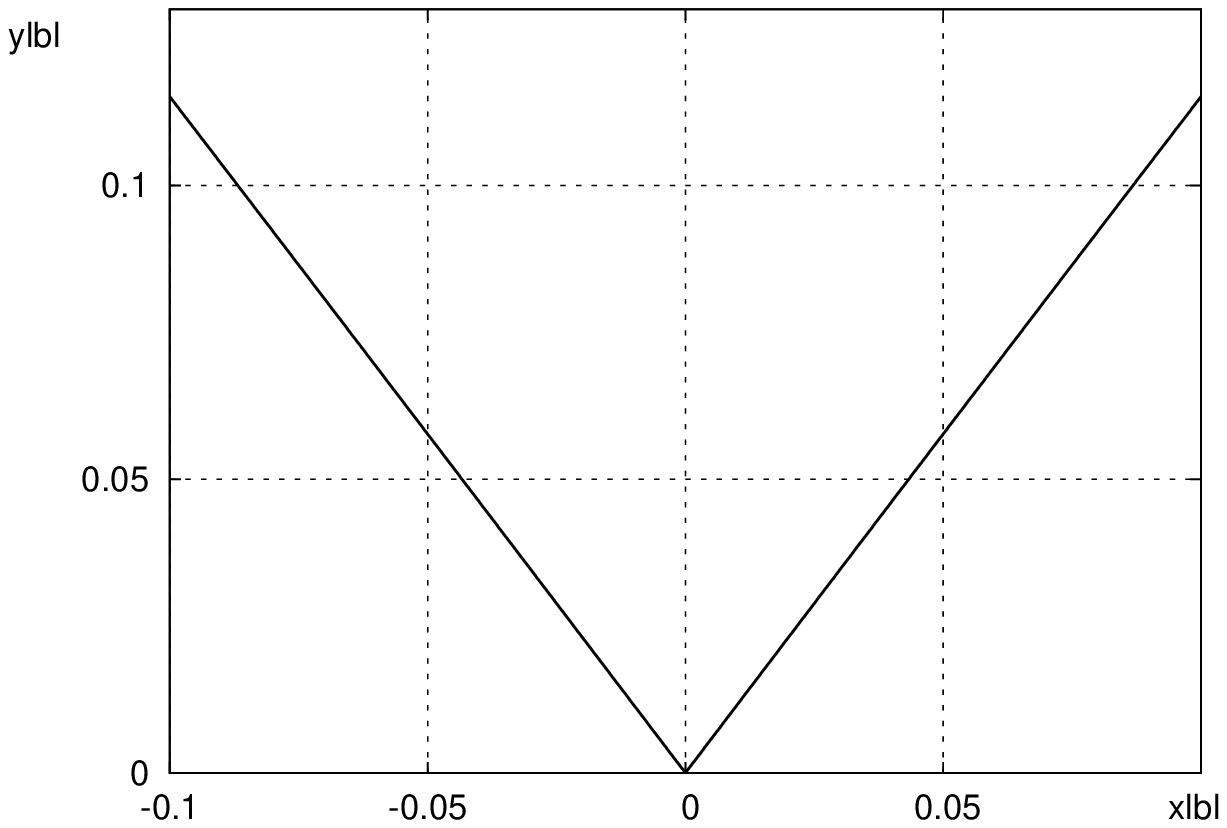}}
\caption{Dependence of $\|R(\sigma)\|$ in \eqref{eq:R_sigma} on $\sigma$ at $k=k_{1}$ near $\sigma_\text{exact}=0$.}
\label{fig:kuramoto_error_0}
\end{figure}

\begin{figure}
\psfrag{xlbl}{$\sigma$}
\psfrag{ylbl}{$\|R(\sigma)\|$}
\center{\includegraphics[width=0.7\linewidth]{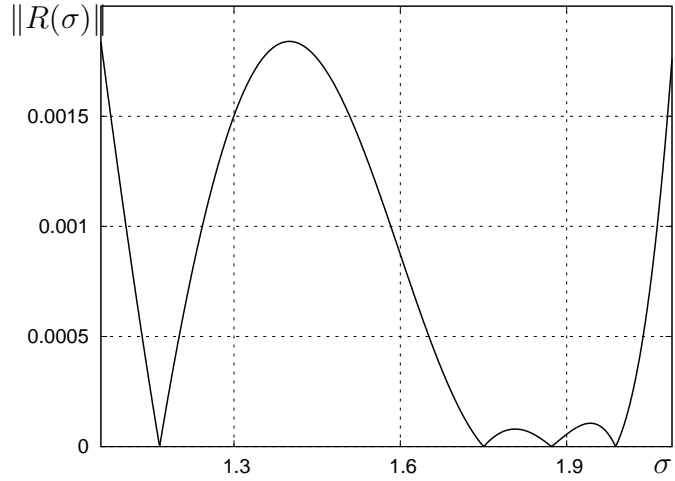}}
\caption{Dependence of $\|R(\sigma)\|$ in \eqref{eq:R_sigma} on $\sigma$ at $k=k_{3}$ near $\sigma_\text{exact}=\frac{12}{\sqrt{47}}, \frac{16}{\sqrt{73}}$.}
\label{fig:kuramoto_error_12sqrt}
\end{figure}

\begin{figure}
\psfrag{xlbl}{$\sigma$}
\psfrag{ylbl}{$\|R(\sigma)\|$}
\center{\includegraphics[width=0.7\linewidth]{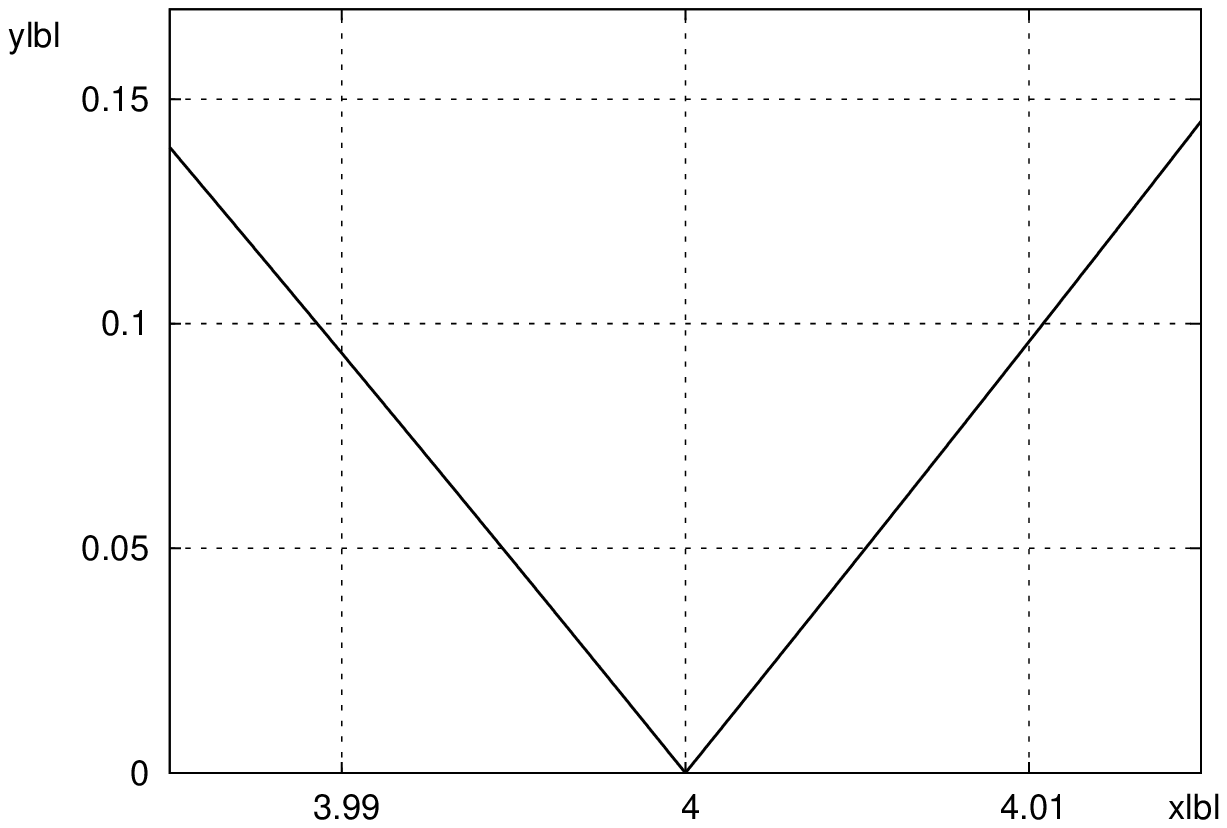}}
\caption{Dependence of $\|R(\sigma)\|$ in \eqref{eq:R_sigma} on $\sigma$ at $k=k_{1}$ near $\sigma_\text{exact}=4$.}
\label{fig:kuramoto_error_4}
\end{figure}

This function has zeros at $\sigma_\text{exact}=0, \frac{12}{\sqrt{47}}, \frac{16}{\sqrt{73}}, 4$.
The dependencies \eqref{eq:R_sigma} on $\sigma$ are presented in Figs. \ref{fig:kuramoto_error_0}, \ref{fig:kuramoto_error_12sqrt} and \ref{fig:kuramoto_error_4} near $\sigma=0$, $\sigma=\frac{12}{\sqrt{47}},\frac{16}{\sqrt{73}}$, $\sigma=4$ respectively.

\begin{figure}
\psfrag{xlbl}{$z$}
\psfrag{ylbl}{$y(z;\sigma$)}
\center{\includegraphics[width=0.7\linewidth]{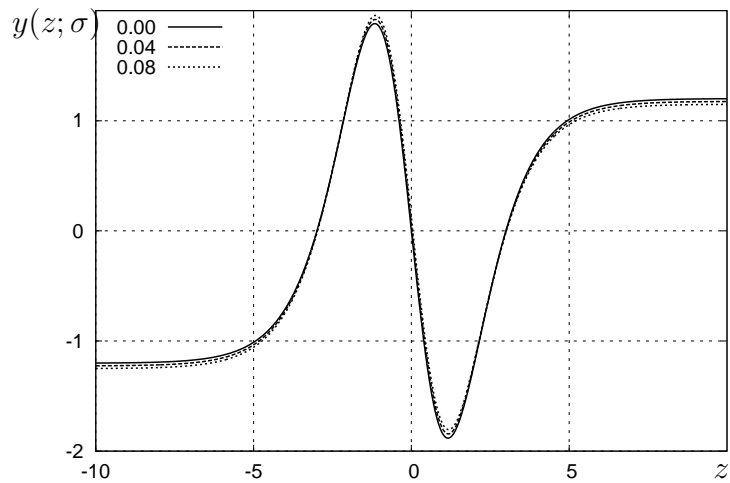}}
\caption{Quasi-exact solution \eqref{eq:kuramoto_sol_a} of Eq.\eqref{Eq1d} at $k=k_{1}$ for $\sigma$: 0.00, 0.04 and 0.08.}
\label{fig:kuramoto_sol_0}
\end{figure}

\begin{figure}
\psfrag{xlbl}{$z$}
\psfrag{ylbl}{$y(z;\sigma$)}
\center{\includegraphics[width=0.7\linewidth]{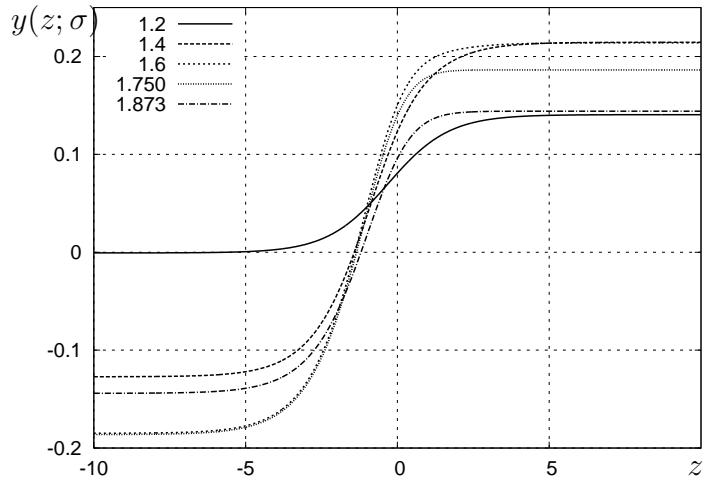}}
\caption{Quasi-exact solution \eqref{eq:kuramoto_sol_a} of Eq.\eqref{Eq1d} at $k=k_{3}$ for $\sigma$: 1.2, 1.4, 1.6, 1.750 and 1.873.}
\label{fig:kuramoto_sol_12sqrt}
\end{figure}

\begin{figure}
\psfrag{xlbl}{$z$}
\psfrag{ylbl}{$y(z;\sigma$)}
\center{\includegraphics[width=0.7\linewidth]{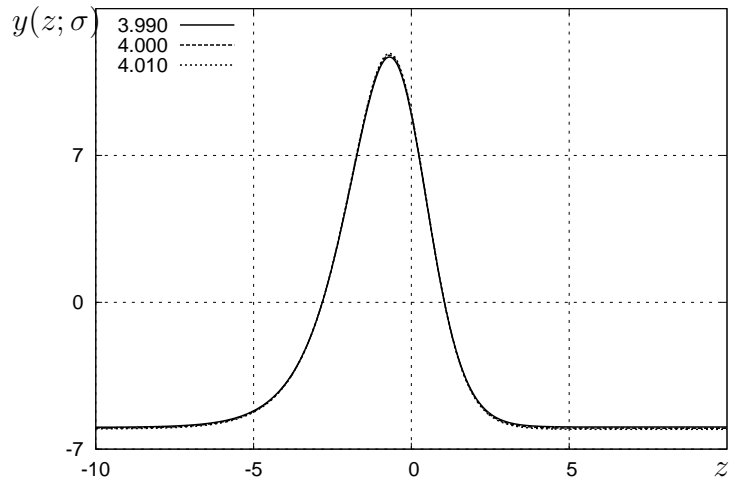}}
\caption{Quasi-exact solution \eqref{eq:kuramoto_sol_a} of Eq.\eqref{Eq1d} at $k=k_{1}$ for $\sigma$: 3.990, 4.000 and 4.010.}
\label{fig:kuramoto_sol_4}
\end{figure}

Results of numerical calculation of values $k$ and $\varepsilon$ are given in Tables \ref{tabl:kuramoto_sigma_range_0}--\ref{tabl:kuramoto_sigma_4} for different values of  $\sigma$.
\begin{table}
    \renewcommand{\arraystretch}{1.2}
    \caption{Values of parameters $\sigma$, $k \simeq k_1$, $a_0$, $a_1$, $a_2$, $a_3$ and error $\varepsilon$ near $\sigma_\text{exact}=0$}
    \label{tabl:kuramoto_sigma_range_0}
\begin{center}
\begin{tabular}{|l|l|l|l|l|l|l|l|}
    \hline
    \multicolumn{1}{|c|}{$\sigma$}
        &\multicolumn{1}{c|}{$k$}
        &\multicolumn{1}{c|}{$a_0$}
        &\multicolumn{1}{c|}{$a_1$}
        &\multicolumn{1}{c|}{$a_2$}
        &\multicolumn{1}{c|}{$a_3$}
        &\multicolumn{1}{c|}{$\varepsilon$} \\
    \hline
    $0.00$      &$0.761$    &$1.201$    &$-28.834$  &$79.292$   &$-52.862$      &$0.000$                \\
    $0.01$      &$0.761$    &$1.195$    &$-28.746$  &$79.204$   &$-52.860$      &$1.154 \cdot 10^{-2}$  \\
    $0.02$      &$0.761$    &$1.189$    &$-28.658$  &$79.112$   &$-52.857$      &$2.309 \cdot 10^{-2}$  \\
    $0.03$      &$0.761$    &$1.182$    &$-28.568$  &$79.017$   &$-52.852$      &$3.463 \cdot 10^{-2}$  \\
    $0.04$      &$0.761$    &$1.176$    &$-28.477$  &$78.919$   &$-52.844$      &$4.616 \cdot 10^{-2}$  \\
    $0.05$      &$0.761$    &$1.169$    &$-28.385$  &$78.818$   &$-52.835$      &$5.768 \cdot 10^{-2}$  \\
    $0.06$      &$0.761$    &$1.163$    &$-28.292$  &$78.714$   &$-52.823$      &$6.921 \cdot 10^{-2}$  \\
    $0.07$      &$0.761$    &$1.156$    &$-28.198$  &$78.606$   &$-52.809$      &$8.070 \cdot 10^{-2}$  \\
    $0.08$      &$0.761$    &$1.150$    &$-28.103$  &$78.496$   &$-52.793$      &$9.219 \cdot 10^{-2}$  \\
    $0.09$      &$0.760$    &$1.143$    &$-28.007$  &$78.382$   &$-52.775$      &$1.037 \cdot 10^{-1}$  \\
    \hline
\end{tabular}
\end{center}
\end{table}
\begin{table}
    \renewcommand{\arraystretch}{1.2}
    \caption{Values of parameters $\sigma$, $k\simeq k_3$, $a_0$, $a_1$, $a_2$, $a_3$ and error $\varepsilon$ near $\sigma_\text{exact}=\frac{12}{\sqrt{47}}, \frac{16}{\sqrt{73}}$}
    \label{tabl:kuramoto_sigma_12sqrt}
\begin{center}
\begin{tabular}{|l|l|l|l|l|l|l|l|}
    \hline
    \multicolumn{1}{|c|}{$\sigma$}
        &\multicolumn{1}{c|}{$k$}
        &\multicolumn{1}{c|}{$a_0$}
        &\multicolumn{1}{c|}{$a_1$}
        &\multicolumn{1}{c|}{$a_2$}
        &\multicolumn{1}{c|}{$a_3$}
        &\multicolumn{1}{c|}{$\varepsilon$} \\
    \hline
    $1.2$       &$0.049$    &$0.141$    &$-0.105$   &$-0.022$   &$-0.014$       &$4.857 \cdot 10^{-4}$  \\
    $1.3$       &$0.096$    &$0.193$    &$-0.145$   &$-0.021$   &$-0.106$       &$1.502 \cdot 10^{-3}$  \\
    $1.4$       &$0.123$    &$0.215$    &$-0.135$   &$0.019$    &$-0.226$       &$1.840 \cdot 10^{-3}$  \\
    $1.5$       &$0.141$    &$0.220$    &$-0.104$   &$0.058$    &$-0.337$       &$1.549 \cdot 10^{-3}$  \\
    $1.6$       &$0.150$    &$0.214$    &$-0.059$   &$0.071$    &$-0.410$       &$8.729 \cdot 10^{-4}$  \\
    $1.7$       &$0.150$    &$0.198$    &$-0.018$   &$0.034$    &$-0.405$       &$2.058 \cdot 10^{-4}$  \\
    $1.750$     &$0.146$    &$0.186$    &$-0.001$   &$-0.002$   &$-0.369$       &$1.090 \cdot 10^{-6}$  \\
    $1.8$       &$0.138$    &$0.172$    &$0.008$    &$-0.043$   &$-0.311$       &$7.858 \cdot 10^{-5}$  \\
    $1.873$     &$0.117$    &$0.144$    &$0.001$    &$-0.096$   &$-0.192$       &$6.947 \cdot 10^{-7}$  \\
    $1.9$       &$0.105$    &$0.131$    &$-0.012$   &$-0.106$   &$-0.140$       &$5.591 \cdot 10^{-5}$  \\
    \hline
\end{tabular}
\end{center}
\end{table}
\begin{table}
    \renewcommand{\arraystretch}{1.2}
    \caption{Values of parameters $\sigma$, $k \simeq k_1$, $a_0$, $a_1$, $a_2$, $a_3$ and error $\varepsilon$ near $\sigma_\text{exact}=4$}
    \label{tabl:kuramoto_sigma_4}
\begin{center}
\begin{tabular}{|l|l|l|l|l|l|l|l|}
    \hline
    \multicolumn{1}{|c|}{$\sigma$}
        &\multicolumn{1}{c|}{$k$}
        &\multicolumn{1}{c|}{$a_0$}
        &\multicolumn{1}{c|}{$a_1$}
        &\multicolumn{1}{c|}{$a_2$}
        &\multicolumn{1}{c|}{$a_3$}
        &\multicolumn{1}{c|}{$\varepsilon$} \\
    \hline
    $3.990$     &$0.998$    &$-5.945$   &$-0.020$   &$119.100$  &$-119.200$     &$9.341 \cdot 10^{-2}$  \\
    $3.995$     &$0.999$    &$-5.972$   &$0.000$    &$119.500$  &$-119.500$     &$4.713 \cdot 10^{-2}$  \\
    $4.000$     &$1.000$    &$-5.999$   &$0.020$    &$120.000$  &$-120.000$     &$0.000$                \\
    $4.005$     &$1.001$    &$-6.028$   &$0.000$    &$120.500$  &$-120.500$     &$4.768 \cdot 10^{-2}$  \\
    $4.010$     &$1.002$    &$-6.054$   &$0.020$    &$120.900$  &$-120.900$     &$9.608 \cdot 10^{-2}$  \\
    \hline
\end{tabular}
\end{center}
\end{table}

Quasi-exact solutions \eqref{eq:kuramoto_sol_a} in the case  $k=k_{1,3}$ with values of the parameters from Tables \ref{tabl:kuramoto_sigma_range_0}--\ref{tabl:kuramoto_sigma_4} are demonstrated in Figs. \ref{fig:kuramoto_sol_0}, \ref{fig:kuramoto_sol_12sqrt} and \ref{fig:kuramoto_sol_4}.

\section{Quasi-exact solutions of the Korteweg--de Vries--Burgers equation}

The Korteweg--de Vries--Burgers equation takes the form
\begin{gather}
    \label{Eq40}
    u_t+u\,u_x+\beta\,u_{xxx}=\nu\,u_{xx},
\end{gather}
where $\nu$ and $\beta$ are constant parameters.

Using the variables
\begin{gather}
    \label{Eq41}
    x=\frac{\nu}{\beta}\,x^{'},\quad t=\frac{\nu^2}{\beta^3}\,t^{'},\quad u = \frac{\beta^2}{\nu}\,u^{'}
\end{gather}
Eq.\eqref{Eq40} can be written as
\begin{gather}
    \label{Eq42}
    u_t+u\,u_x+\,u_{xxx}=\,u_{xx},
\end{gather}
where the primes are omitted.

Seeking traveling wave solutions
\begin{gather}
    \label{Eq43}
    u(x,t)=y(z), \quad z=k\,x-\omega\,t,
\end{gather}
after one integration we have the nonlinear ODE
\begin{gather}
    \label{Eq44}
    k^3\,y_{zz}-k^2\,y_z+\frac12\,k\,y^2-\omega\,y+C_1=0, \quad k \neq 0.
\end{gather}
Using a new variable $y(z)=y'(z)+\frac{\omega}{k}$ (the primes are omitted) we transform Eq.\eqref{Eq44} into the form
\begin{gather}
    \label{Eq442}
    k^3\,y_{zz}-k^2\,y_z+\frac12\,k\,y^2-\frac{\omega^2}{2k}+C_1=0, \quad k \neq 0.
\end{gather}
Using the substitution $k \rightarrow -k, z \rightarrow -z, C_1 \rightarrow -C_1$ we have the same equation, so we focus on the case $k > 0$.
One can verify that the general solution of Eq.\eqref{Eq442} has a pole of the second order.
Hence, we look for exact solution of Eq.\eqref{Eq442} in the form
\begin{gather}
    \label{Eq45}
    y(z)=a_0+a_1\,Q(z)+a_2\,Q(z)^2,
\end{gather}
where $Q(z)$ is determined by formula \eqref{Eq10d}.

Substituting Eq.\eqref{Eq45} into Eq.\eqref{Eq442} leads to the following system of algebraic equations:
\begin{equation}
\label{eq:kdvb_a_sys}
\begin{split}
&Q^4: a_2 + 12 k^2 = 0, \\
&Q^3: 2 k^2 a_1 - 10 k^2 a_2 + a_1 a_2 - 2 k a_2 = 0, \\
&Q^2: - 4 k a_2 + 2 k a_1 - a_1^2 + 6 k^2 a_1 - 8 k^2 a_2 - 2 a_0 a_2 = 0, \\
&Q^1: a_1 \left( k^2 + k + a_0 \right) = 0, \\
&Q^0: 2 C_1 k + k^2 a_0^2 - \omega^2 = 0,
\end{split}
\end{equation}
with exact solution
\begin{equation}
\label{Eq46}
a_2 = -12\,k^2, \quad
a_1 = 12\,k^2+\frac{12}{5}\,k,\quad
a_0 = -k^2-\frac65\,k+\frac{1}{25}, \quad
C_1 = \frac{\omega^2}{2k} - \frac{k a_0^2}{2}.
\end{equation}
From system \eqref{eq:kdvb_a_sys} we also have an equation for the determination of $k$
\begin{equation}
\label{eq:kdvb_k_eq}
25 k^2 - 1 = 0,
\end{equation}
yielding
\begin{equation}
k_\text{exact} = \frac{1}{5}.
\end{equation}

Taking \eqref{Eq46} into account, the quasi-exact solution \eqref{Eq45} takes the form
\begin{equation}
\label{eq:kdvb_sol_a}
y(z;k) = \frac{1}{25}- \frac{6}{5} k  - k^2 + \frac{12}{5} k ( 5 k + 1 ) Q(z) - 12 k^2 Q(z)^2.
\end{equation}

\begin{figure}
\psfrag{xlbl}{$k$}
\psfrag{ylbl}{$\|R(k)\|$}
\center{\includegraphics[width=0.7\linewidth]{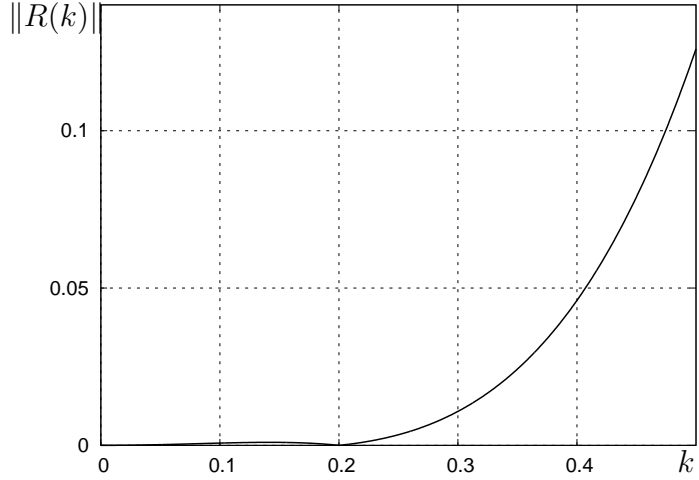}}
\caption{Dependence of $\|R(k)\|$ in \eqref{eq:kdvb_error_norm} near $k_\text{exact}=\frac{1}{5}$.}
\label{fig:kdvb_res}
\end{figure}

The deviation function for solution \eqref{eq:kdvb_sol_a} at the condition $C_1=\frac{\omega^2}{2k} - \frac{k a_0^2}{2}$ has the form
\begin{equation}
R(z;k) = \frac{12}{125} k^2 (25 k^2 - 1) Q(z)
\label{eq:kdvb_error}
\end{equation}
with norm
\begin{equation}
\|R(k)\| = \frac{12}{125} k^2 |25 k^2 - 1|.
\label{eq:kdvb_error_norm}
\end{equation}
The dependence of the norm of the deviation function on parameter $k$ is given in Fig. \ref{fig:kdvb_res}.

The values of parameters $k$, $a_0$, $a_1$, $a_2$ and error $\varepsilon=\|R(\sigma)\|$ are given in Table \ref{tabl:kdvb_k_0.2}.
\begin{table}
    \renewcommand{\arraystretch}{1.2}
    \caption{Values of parameters $a_0$, $a_1$, $a_2$, and error $\varepsilon$ near $k_\text{exact}=\frac{1}{5}$.}
    \label{tabl:kdvb_k_0.2}
\begin{center}
\begin{tabular}{|l|l|l|l|l|l|}
    \hline
    \multicolumn{1}{|c|}{$k$}
        &\multicolumn{1}{c|}{$a_0$}
        &\multicolumn{1}{c|}{$a_1$}
        &\multicolumn{1}{c|}{$a_2$}
        &\multicolumn{1}{c|}{$\varepsilon$} \\
    \hline
        $0.05$      &$0.098$    &$-0.090$   &$-0.030$   &$2.250 \cdot 10^{-04}$     \\
        $0.1$       &$0.150$    &$-0.120$   &$-0.120$   &$7.200 \cdot 10^{-04}$     \\
        $0.15$      &$0.198$    &$-0.090$   &$-0.270$   &$9.450 \cdot 10^{-04}$     \\
        $0.2$       &$0.240$    &$0.000$    &$-0.480$   &$0.000$                    \\
        $0.25$      &$0.278$    &$0.150$    &$-0.750$   &$3.375 \cdot 10^{-03}$     \\
        $0.3$       &$0.310$    &$0.360$    &$-1.080$   &$1.080 \cdot 10^{-02}$     \\
        $0.35$      &$0.338$    &$0.630$    &$-1.470$   &$2.426 \cdot 10^{-02}$     \\
        $0.4$       &$0.360$    &$0.960$    &$-1.920$   &$4.608 \cdot 10^{-02}$     \\
        $0.45$      &$0.378$    &$1.350$    &$-2.430$   &$7.898 \cdot 10^{-02}$     \\
        $0.5$       &$0.390$    &$1.800$    &$-3.000$   &$1.260 \cdot 10^{-01}$     \\
    \hline
\end{tabular}
\end{center}
\end{table}

\begin{figure}
\psfrag{xlbl}{$z$}
\psfrag{ylbl}{$y(z;k)$}
\center{\includegraphics[width=0.7\linewidth]{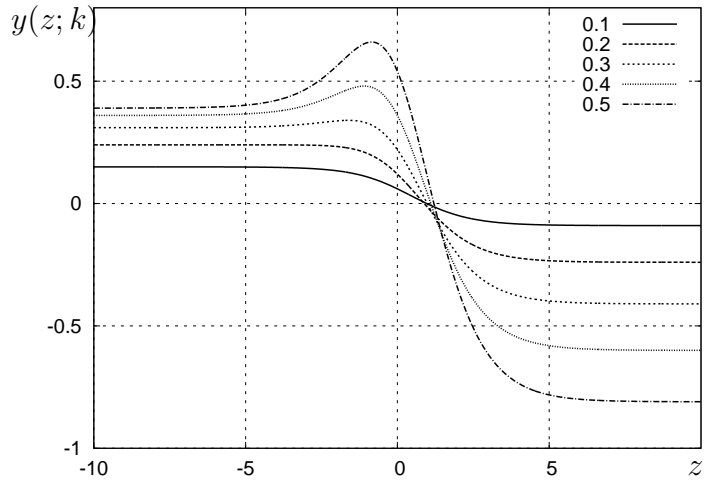}}
\caption{Quasi-exact solution \eqref{Eq45} for values of $k$: 0.1, 0.2, 0.3, 0.4 and 0.5}
\label{fig:kdvb_sol}
\end{figure}

The quasi-exact solutions \eqref{eq:kdvb_sol_a} with some values of parameters from Table \ref{tabl:kdvb_k_0.2} are shown in Fig. \ref{fig:kdvb_sol}.

\section{Quasi-exact solutions of the Kawahara equation}

The Kawahara equation \cite{Kawahara72} takes the form
\begin{equation}
u_t + uu_x + u_{xxx} = u_{xxxxx}.
\label{eq:kawahara_partial}
\end{equation}

Seeking traveling wave solutions,
\begin{equation}
u(x,t) = y(z), \quad z = k x - \omega t,
\end{equation}
we have  the nonlinear ODE (after one integration with respect to $z$)
\begin{equation}
C_1 - \omega y + \frac{1}{2} ky^2 + k^3 y_{zz} - k^5 y_{zzzz} = 0, \quad k \neq 0.
\label{eq:kawahara_ode}
\end{equation}

Using the new variable $y(z) = \frac{\omega}{k} + y'(z)$ (the primes are omitted) Eq.\eqref{eq:kawahara_ode} can be transformed into
\begin{equation}
C_1 - \frac{\omega^2}{2k} + \frac{1}{2} ky^2 + k^3 y_{zz} - k^5 y_{zzzz} = 0, \quad k \neq 0.
\label{eq:kawahara}
\end{equation}
Eq.\eqref{eq:kawahara} is invariant when $k \rightarrow -k$, $C_1\rightarrow -C_1$, so, we consider the case $k > 0$.
The general solution has a pole of fourth order.
Hence, we look for quasi-exact solutions of Eq.\eqref{eq:kawahara} in the form
\begin{equation}
y(z)=\sum_{i=0}^4 a_i Q(z)^i.
\label{eq:kawahara_y_Q}
\end{equation}

Substituting \eqref{eq:kawahara_y_Q} into Eq.\eqref{eq:kawahara} and equating expressions at different powers of $Q(z)$ we obtain the system of algebraic equations for the parameters $a_4$, $a_3$, $a_2$, $a_1$, $k$ and $C_1$
\begin{equation}
\label{eq:kawahara_a_sys}
\begin{split}
&Q^8: a_4 - 1680 k^4 = 0, \\
&Q^7: - a_3 a_4 + 360 k^4 a_3 - 2640 k^4 a_4 = 0, \\
&Q^6: 240 k^4 a_2 + 6040 k^4 a_4 - a_3^2 - 2 a_2 a_4 - 40 k^2 a_4 - 2160 k^4 a_3 = 0, \\
&Q^5: 36 k^2 a_4 - a_1 a_4 - 1476 k^4 a_4 - 12 k^2 a_3 - 336 k^4 a_2 + 1164 k^4 a_3 + 24 k^4 a_1 - a_2 a_3 = 0, \\
&\begin{split}
Q^4: 12 k^2 a_2 - 42 k^2 a_3 + 32 k^2 a_4 + 2 a_0 a_4 - 660 k^4 a_2 &+ 2 a_1 a_3 + 1050 k^4 a_3 \\
&+ a_2^2 - 512 k^4 a_4 + 120 k^4 a_1 = 0,
\end{split} \\
&Q^3: 81 k^4 a_3 + 50 k^4 a_1 - a_0 a_3 - 9 k^2 a_3 - 2 k^2 a_1 - a_1 a_2 - 130 k^4 a_2 + 10 k^2 a_2 = 0, \\
&Q^2: 8 k^2 a_2 + a_1^2 + 2 a_0 a_2 - 6 k^2 a_1 + 30 k^4 a_1 - 32 k^4 a_2 = 0, \\
&Q^1: a_1 ( - a_0 + k^4 - k^2 ) = 0, \\
&Q^0: 2 C_1 k + k^2 a_0^2 - \omega^2 = 0.
\end{split}
\end{equation}

Solving system \eqref{eq:kawahara_a_sys},
\begin{equation}
\begin{split}
a_4 &= 1680 k^4, \quad
a_3 = - 3360 k^4, \quad
a_2 = 1960 k^4 - \frac{280 k^2}{13}, \\
a_1 &= - 280 k^4 + \frac{280 k^2}{13}, \quad
a_0 = - \frac{7 k^4}{3} - \frac{70 k^2}{39} - \frac{31}{507}, \quad
C_1 = \frac{\omega^2}{2k} - \frac{k a_0^2}{2}.
\end{split}
\label{eq:kawahara_a_exact}
\end{equation}
together with the equation that determines $k$,
\begin{equation*}
( 13 k^2 - 1 ) ( 1690 k^4 + 403 k^2 + 31 ) = 0.
\end{equation*}
The real solution is
\begin{equation*}
k_\text{exact} = \frac{1}{\sqrt{13}}.
\end{equation*}

Quasi-exact solution \eqref{eq:kawahara_y_Q} of Eq.\eqref{eq:kawahara} then takes the form
\begin{equation}
\begin{aligned}
y(z) = &- \frac{7 k^4}{3} - \frac{70 k^2}{39} - \frac{31}{507} - \frac{280 k^2}{13} ( 13 k^2 - 1 ) Q(z) \\
&+ \frac{280 k^2}{13} ( 91 k^2 - 1 ) Q(z)^2 - 3360 k^4 Q(z)^3 + 1680 k^4 Q(z)^4.
\end{aligned}
\label{eq:kawahara_y_Q_a}
\end{equation}

\begin{figure}
\psfrag{xlbl}{$k$}
\psfrag{ylbl}{$\|R(k)\|$}
\center{\includegraphics[width=0.7\linewidth]{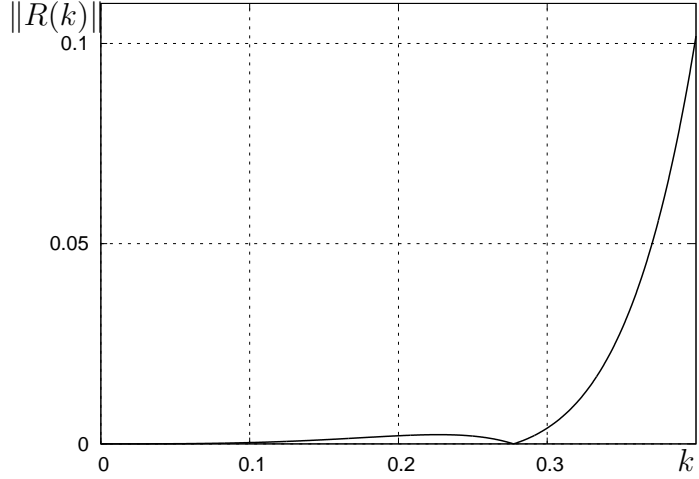}}
\caption{Dependence of $\|R(k)\|$ in \eqref{eq:kawahara_error_norm} near $k=\frac{1}{\sqrt{13}}$.}
\label{fig:kawahara_error_norm}
\end{figure}

Substituting \eqref{eq:kawahara_y_Q_a} into Eq.\eqref{eq:kawahara} we obtain the deviation function in the form
\begin{equation}
R(z;k) = - \frac{280}{6591} ( 1690 k^4 + 403 k^2 + 31 ) k^3 ( 13 k^2 - 1 ) \left( Q(z)^2 - Q(z) \right),
\label{eq:kawahara_error}
\end{equation}
with norm
\begin{equation}
\|R(k)\| = \frac{70}{6591} \; ( 1690 k^4 + 403 k^2 + 31 ) \left| k^3 ( 13 k^2 - 1 ) \right|,
\label{eq:kawahara_error_norm}
\end{equation}
which equals zero at $k_\text{exact}=\frac{1}{\sqrt{13}}$.
The dependence of \eqref{eq:kawahara_error_norm} on $k$ near $k=\frac{1}{\sqrt{13}}$ is presented in Fig. \ref{fig:kawahara_error_norm}.

The values $a_0$, $a_1$, $a_2$, $a_3$, $a_4$ and $\varepsilon=\|R(k)\|$ for different values of $k$ near $\frac{1}{\sqrt{13}}$ are given in Table \ref{tabl:kawahara_k_1sqrt13}.
\begin{table}
    \renewcommand{\arraystretch}{1.2}
    \caption{Values of parameters $a_0$, $a_1$, $a_2$, $a_3$, $a_4$, and error $\varepsilon$ near $k_\text{exact}=\frac{1}{\sqrt{13}}$}
    \label{tabl:kawahara_k_1sqrt13}
\begin{center}
\begin{tabular}{|l|l|l|l|l|l|l|l|}
    \hline
    \multicolumn{1}{|c|}{$k$}
        &\multicolumn{1}{c|}{$a_0$}
        &\multicolumn{1}{c|}{$a_1$}
        &\multicolumn{1}{c|}{$a_2$}
        &\multicolumn{1}{c|}{$a_3$}
        &\multicolumn{1}{c|}{$a_4$}
        &\multicolumn{1}{c|}{$\varepsilon$} \\
    \hline
        $0.05$      &$-0.066$   &$0.052$    &$-0.042$   &$-0.021$   &$0.011$    &$4.112 \cdot 10^{-05}$     \\
        $0.1$       &$-0.079$   &$0.187$    &$-0.019$   &$-0.336$   &$0.168$    &$3.252 \cdot 10^{-04}$     \\
        $0.15$      &$-0.103$   &$0.343$    &$0.508$    &$-1.701$   &$0.851$    &$1.038 \cdot 10^{-03}$     \\
        $0.2$       &$-0.137$   &$0.414$    &$2.274$    &$-5.376$   &$2.688$    &$2.032 \cdot 10^{-03}$     \\
        $0.25$      &$-0.182$   &$0.252$    &$6.310$    &$-13.125$  &$6.563$    &$1.954 \cdot 10^{-03}$     \\
        $0.277$     &$-0.213$   &$0.000$    &$9.941$    &$-19.882$  &$9.941$    &$4.093 \cdot 10^{-05}$     \\
        $0.3$       &$-0.242$   &$-0.330$   &$13.938$   &$-27.216$  &$13.608$   &$3.947 \cdot 10^{-03}$     \\
        $0.35$      &$-0.316$   &$-1.563$   &$26.774$   &$-50.421$  &$25.211$   &$2.853 \cdot 10^{-02}$     \\
        $0.4$       &$-0.408$   &$-3.722$   &$46.730$   &$-86.016$  &$43.008$   &$1.019 \cdot 10^{-01}$     \\
    \hline
\end{tabular}
\end{center}
\end{table}

\begin{figure}
\psfrag{xlbl}{$z$}
\psfrag{ylbl}{$y(z;k)$}
\center{\includegraphics[width=0.7\linewidth]{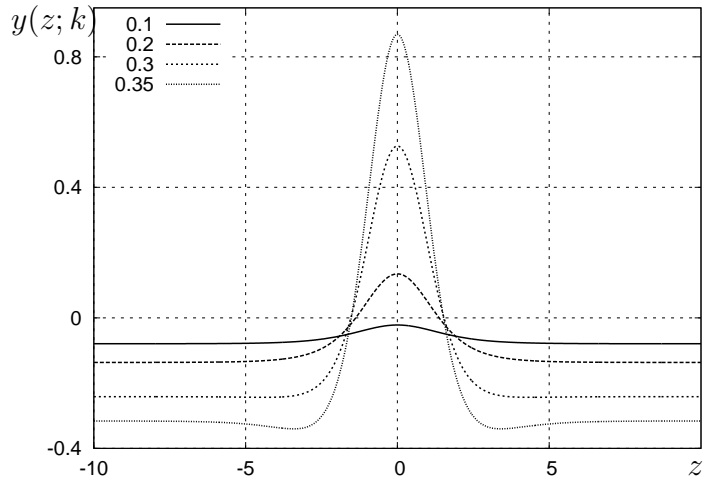}}
\caption{Quasi-exact solution \eqref{eq:kawahara_y_Q_a} for values of $k$: 0.1, 0.2, 0.3 and 0.35.}
\label{fig:kawahara_sol}
\end{figure}

Quasi-exact solutions \eqref{eq:kawahara_y_Q_a} with the values of the parameters from Table \ref{tabl:kawahara_k_1sqrt13} are presented in Fig. \ref{fig:kawahara_sol}.

\section{Conclusion}
In this paper we have introduced the concept of the quasi-exact solutions of nonlinear differential equations.
These solution are close to exact solutions but after substitution into differential equation give a small error.
We have found quasi-exact solutions for the Kuramoto--Sivashinsky, the Korteweg--de Vries--Burgers, and the Kawahara equations.
In comparison with exact solutions the quasi-exact solutions can be used for the description of physical processes of the approximate mathematical models.

\section{Acknowledgements}
This research was partially supported by Federal Target Programmes ``Research and Scientific---Pedagogical Personnel of Innovation in Russian Federation on 2009-–2013'' and ``Research and developments in priority directions of development of a scientifically-technological complex of Russia on 2007--2013''.
One of authors (N.A. Kudryashov) thanks professor W. Hereman for useful remarks.

\end{document}